\documentclass[11pt]{article}
\usepackage{moriond,epsfig}

\def\be{\begin{equation}}
\def\ee{\end{equation}}
\def\bea{\begin{eqnarray}}
\def\eea{\end{eqnarray}}

\begin{document}
\vspace*{4cm}
\title{THE BIAS OF GALAXIES AND THE DENSITY OF THE UNIVERSE FROM THE 2dF
GALAXY REDSHIFT SURVEY}

\author{Licia Verde$^{1,2}$, Alan Heavens$^{3}$, Will Percival$^{3}$, Sabino Matarrese$^{4}$}

\address{$^{1}$Department of Physics and Astronomy, Rutgers University, 136
Frelinghuysen Road, Piscataway NJ 08854-8019, USA, $^{2}$Dept. of  Astrophysical
Sciences, Princeton University, Princeton NJ 08540-1001, USA, $^{3}$Institute
for Astronomy, Blackford Hill, Edinburgh (UK), $^{4}$ Dipartimento di Fisica
Galileo Galilei, Universit\'{a} di Padova, Padova (IT)}

\maketitle\abstracts{ By studying the bispectrum of the galaxy distribution in the 2dF
galaxy redhsift survey (2dFGRS), we have shown that 2dFGRS galaxies are unbiased tracers of the mass
distribution. This allows us the break the degeneracy intrinsic to power
spectrum studies, between the matter density parameter $\Omega_m$ and the bias
parameter $b$, and to obtain an accurate  measurement of
$\Omega_m$: $\Omega_{m,z_{eff}}=0.27 \pm 0.06$,  a measurement obtained from  the
2dFGRS  alone, independently from other data sets.
This result has to be interpreted as
$\Omega_m$ at the effective redshift of the survey $z_{eff}=0.17$ Extrapolated
at $z=0$ we obtain $\Omega_m=0.2 \pm 0.06$ for the $\Lambda$CDM model.
This constraint on the matter density parameter when combined with cosmic microwave background constraints on the
flatness of the Universe, show firm evidence that the Universe is dominated by
a vacuum energy component.}

\section{What's bias}
We are confident that the Universe started off almost uniform with very small
primordial perturbations as we can see for example in the Cosmic microwave
background at redshift $\sim 1100$, with a  distribution very close to
gaussian (cf. Heavens, Wu and Sanz contributions).
Given these properties of the inital conditions and a set of cosmological
parameters, it is
possible to predict the statistical properties of the large scale mass
distribution in the local Universe (at $z \sim 0$) via e.g. N-body simulations.
 If we could observe the
mass distribution in the local Universe this would be a very powerful tool
to/we could  discriminate between different cosmological models.
Unfortunately, we cannot observe the mass distribution directly, what we can
easily observe is the  distribution of objects that ``light up'' as for
example the galaxy distribution  in the 2dF galaxy
redshift survey (2dFGRS; fig. 1 left panel).
It is well known that different kind of luminous objects show different
clustering properties (e.g., Peacock and Dodds 1994). Thus different kind of
luminous objects cannot be all faithful tracers of the underlying mass
distrbution, and galaxies might be biased tracers of the mass: on large
scales we might say $\delta_g =
b \delta_m$, where $\delta\equiv\delta \rho/\rho$ the subscripts $g$ and $m$
stand for galaxy and mass, and $b$ denotes the (linear) bias parameter. The concept of
bias was introduced by Kaiser (1984), originally to explain the clustering
properties of Abell clusters, but the  concept was a key element to
reconcile the theoretical prejudice of an $\Omega_m=1$ Universe\footnote{
$\Omega_m$ denotes the present-day matter density of the Universe in units of the critical density.}  with the
observations of large-scale structure. Today we know that the Universe is not
Einstein-de Sitter, so in principle we do not need galaxy bias, but 'the genie
is out of the bottle'.  
\subsection{Why measure it}
We could say that bias encloses our ignorance about the  complicated
process of galaxy formation, so we could learn about galaxy formation by
measuring it.  However, there is a more important reason to measure it:  there is a degeneracy
intrinsic to linear theory  large-scale structure studies (i.e. power
spectrum or correlation function studies) between the effect of gravity driven
by $\Omega_m$ and the effect of bias; these studies can only yield
$\beta\simeq \Omega_m^{0.6}/b$, thus we need to know, or measure,  the bias
parameter $b$ to get the real ``prize'', $\Omega_m$.
 The 2dF team has measured this parameter from the survey
obtaining $\beta=0.43 \pm 0.07$ (Peacock et al. 2001).

\subsection{How to disentangle gravity from bias}
Gravity leaves its own signature on the distribution of mass density.
In fact non-linear gravity skews the field (because of the requirement that  $\delta_m \ge -1$). Bias can also
introduce skewness, but the effect of gravity on the {\it shape} of the
cosmological structures is different from that of biasing: gravity creates
very characteristic  sheets and filaments in the mass distribution
(e.g. Zeldovich pancakes).
If we succed in  measuring how much gravity signature there is in the galaxy
distribution, we can disentangle gravity from biasing. 
The statistical tool most effective to pick up this signature of gravity is the bispectrum.


\section{Bispectrum}
The bispectrum, $B(\vec{k_1}, \vec{k_2}, \vec{k_3})$, is the Fourier space counterpart of the three point
correlation function, more precisely: $\langle
\delta_{\vec{k_1}}\delta_{\vec{k_2}}\delta_{\vec{k_2}} \rangle=B(\vec{k_1},
\vec{k_2}, \vec{k_3})\delta^D(\vec{k_1}+\vec{k_2}+ \vec{k_3})$ where $\delta^D$
denotes the Dirac delta function.
Due to the presence of the Dirac delta function we see that the bispectrum is
defined on triplets of k-vectors that form  triangles.    
The bispectrum is zero for a gaussian field, but, even if the Universe started
off with gaussian initial conditions, we should expect to detect non-zero
bispectrum today on non-linear scales. In particular, the expression for the
bispectrum is easy to obtain in the mildly non-linear regime i.e. in second
order perturbation theory in $\delta$. (The theory of the bispectrum has been
developed by e.g., Fry 1994,
Matarrese Verde Heavens 1997, Verde et al 1998, Scoccimarro et al 1998,  Scoccimarro et al 1999). To be
consistent one has to expand  also the bias to second order
(i.e. $\delta_g=b_1\delta+m+1/2b_2\delta_m^2+...$).
The expression for the galaxy bispectrum thus becomes:
\begin{equation}
B_g(\vec{k_1}, \vec{k_2}, \vec{k_3})=\left[\frac{1}{b_1} J(\vec{k_1}, \vec{k_2})+\frac{b_2}{b_1^2}\right]P_g(k_1)P_g(k_2)+cyc.
\end{equation} 
where $P_g$ denotes the (measurable) galaxy power spectrum, $B_g$ the
(measurable) galaxy bispectrum and $J$ is a known function of
the k-vectors. The shape information, that is the signature of gravity in the
large scale structure distribution is enclosed in this function $J$.   

The form of equation  (1) implies that it is possible  to extract the linear and quadratic bias
parameters ($b_1$ and $b_2$) via a likelihood analysis.

\section{The bias of 2dF galaxies}
 The 2dFGRS, when finished,  will provide a 3D map of
250,000 galaxies\footnote{for more details see Lahav contribution in
these proceedngs or  {\sf http://www.mso.anu.edu.au/2dFGRS/}}.
The bispectrum analysis presented here is based on only 130K of them. 
Equation (1) applies only on mildly non-linear
scales, and, more importantly, only in a very idealized case (survey of
infinite volume, the distance of galaxies is accurately known, no discreteness effects etc.).
In the case of the 2dFGRS
there are many real-world effects to account for (redshift space distortions,
discreteness, window function, selection function etc..). We can model all
these effects  (for details see Verde et al. 1998, Verde et al. 2002)  and
asses the performance of the bispectrum method with a  Monte Carlo
approach, by performing the analysis on several mock 2dFGRS catalogs with known
bias parameters. 

We analized the bispectrum signal of  80 million triangles configurations from
the galaxy distribution of the 2dFGRS  obtaining: $b_1=1.04 \pm 0.11$,
$b_2=-0.05\pm 0.08$. The 2dF galaxies on large
scales are thus  unbiased tracers of the  mass distribution. 
This result is in
agreement with the findings of Lahav et al. 2002 (see Lahav contributon), but
has been obtained with a completely independent method and entirely from the
2dFGRS alone, independently of other data sets. 
The fact that the quadratic bias parameter is consistent with being zero
argues powerfully that 2dFGRS galaxies indeed trace the mass on large scales.

We also find no evidence for scale dependence bias. At some level
non-linear/scale dependent bias must appear on small scales, but these scales might
be too small for the perturbative method used here to be valid.
Note that this results applyes to 2dF galaxies or more specifically to $1.9
L_*$ galaxies at $z \sim 0.17$. Because of the tendency of bias to approach
unity with time,  this results does not rule out  the biased
galaxy formation picture: galaxies might have had a significant bias at
formation time.  

\section{The density of the Universe}
By combining the 2dFGRS measurement of the $\beta $ parameter with the
measurement of the bias presented here, we obtain a determination of the matter
density of the Universe $\Omega_{m,z_{eff}}=0.27 \pm 0.06$. 
We stress here that this result has been obtained from the 2dFGRS alone,
independently of other data sets.
Note that the effective
depth of the survey is $z=0.17$, thus our measurement should be interpreted as
$\Omega_m$ at this epoch.  The extrapolation at $z=0$ is model dependent. In
the right panel of fig. 1  the blue shaded areas show the 1,2 and 3 sigma confidence contours in the $\Omega_m$,
$\Omega_{\Lambda}$ plane (here $\Lambda$ denotes the cosmological constant).
For comparison we also show the 1,2 and 3 sigma constraints from the Boomerang data
(de Bernardis 2001) and supernovae 1A (Perlmutter et al 1999). The dotted line
denotes flat Universe.
The solid lines show the joint confidence contours for Boomerang and 2dF. 
It is quite remarkable that three different experiments (subject to very
dfferent systematics) agree so well (Heavens, Verde, Percival 2002) and point
towards the same cosmological model.

\section{Conclusions}
We have demonstrated that the optically selected galaxies from the  2dFGRS
trace the dark matter density extremely well on large scales.
This allows us to break the degeneracy, intrinsic
to power spectrum studies,  between gravity and biasing. We obtain a measurement of the matter density
parameter at $z_{eff}=0.17$ of $\Omega_{m,z_{eff}}=0.27 \pm 0.06$. The extrapolation of this
quantity at $z=0$ is model dependent, however the constraints of cosmological
parameters obtained from 2dFGRS, Cosmic microwave background and
Supernovae agree remarkably well. Thus we can cosnculd that we  are converging towards a cosmological
model where the Universe is flat, is dominated by a vacuum energy component
($\Lambda=0.8 \pm 0.08$) and the matter density is $\Omega_m=0.2 =\pm 0.06$.


\begin{figure}
\psfig{figure=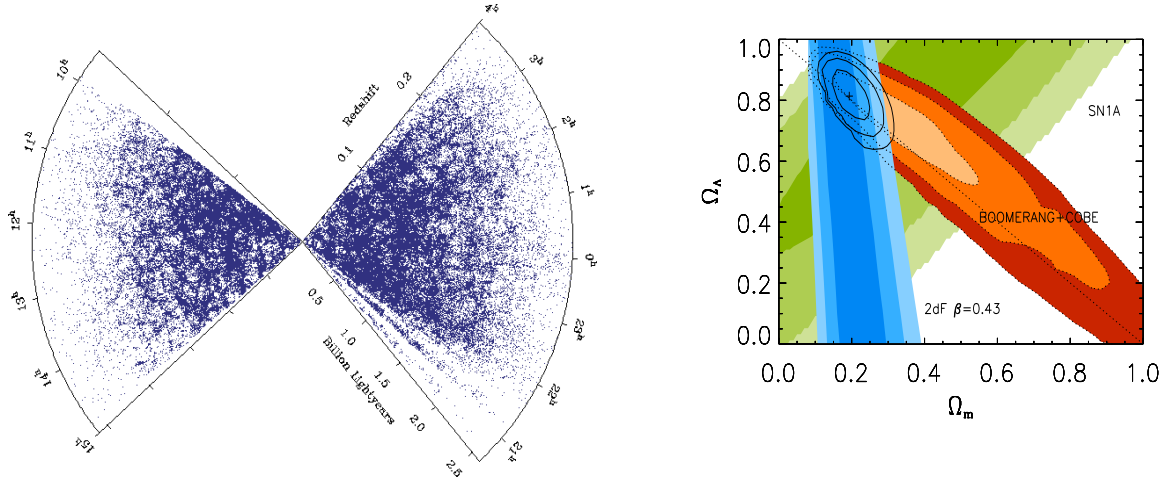,height=2.5in}
\caption{LEFT: the 2dF galaxy redshift survey. RIGHT:1,2 and 3 sigma confidence contours in the $\Omega_m$, $\Omega_{\Lambda}$
plane: green are the supernovae 1A consraints (Perlmutter et al 1999), red are
the  CMB constraints from Boomerang (de Bernardis et al. 2000), blue are the LSS constraints
obtained by the present work (Verde et al 2002). The solid lines show the
joint likelihood contours of CMB +LSS data sets.
It is truly remarkable that different measuerement of cosmological paraneters that
use different observables and  are subject to different systematics agree so well.
\label{fig:cmblsssn}}
\end{figure}

\section*{Acknowledgments}
We would like to thank the 2dFGRS team and  AAO staff for making this superb galaxy survey.


\begin{thebibliography}{99}
\bibitem{} de Bernardis et al.  2000, Nature, 404, 955
\bibitem{} Fry, J. 1994, Phys. Rev. Lett., 73, 215
\bibitem{} Heavens A. F., Verde, L., Percival W. 2002, submitted.
\bibitem{} Kaiser, N. 1984, ApJ(Lett), 284, L9
\bibitem{} Lahav et al. 2002, MNRAS in press (astro-ph/0112162)
\bibitem{} Matarrese, S., Verde L.,  Heavens A. F. 1997, MNRAS, 290, 651
\bibitem{} Peacock et al. 2001, Nature, 410, 169
\bibitem{} Perlmutter et al. 1999, ApJ, 483, 565
\bibitem{} Scoccimarro et al., 1998, ApJ, 496, 586
\bibitem{} Scoccimarro R., Couchman H., Frieman J. A. 1999, ApJ, 517, 531
\bibitem{} Verde et al. 1998, MNRAS, 300, 747
\bibitem{} Verde et al. 2002, MNRAS in press (astro-ph/0112161)


\end{thebibliography}
\end{document}